\pdfoutput=1

\documentclass[times,twocolumn,final,authoryear]{elsarticle}

\usepackage{jasr}
\usepackage{framed,multirow}
\usepackage[yyyymmdd,hhmmss]{datetime}
\usepackage{amssymb}
\usepackage{latexsym}
\usepackage{siunitx}
\DeclareSIUnit{\parsec}{\text{\ensuremath{\mathrm{pc}}}}
\DeclareSIUnit{\angstrom}{\textup{\AA}}

\usepackage{url}
\usepackage{xcolor}
\definecolor{newcolor}{rgb}{.8,.349,.1}

\usepackage[
breaklinks=true,%
colorlinks=true,%
pdfauthor={Neilsen},%
pdftitle={Massive, Evolving Winds in Black Hole X-ray Binaries}%
]{hyperref}
\usepackage[noabbrev]{cleveref}
\usepackage{caption}
\usepackage{subcaption}
\usepackage{aas_macros}

\journal{Advances in Space Research}

\begin{document}

\verso{G. Shaifullah \textit{et. al}}

\begin{frontmatter}


\title{Validation of heliospheric modeling algorithms through pulsar observations II: simulations with EUHFORIA }

\author[1,2]{G. M. Shaifullah\corref{cor}}
\ead{golam.shaifullah@unimib.it}
\cortext[cor]{Corresponding author}

\author[3,4]{J. Magdalenic}
\author[5,6]{C. Tiburzi}
\author[3,4]{I. Jebaraj}
\author[3,4]{E. Samara}
\author[6]{P. Zucca}

\address[1]{Dipartimento di Fisica ``G. Occhialini'', Universit\`a di Milano-Bicocca, Piazza della Scienza 3, 20126 Milano, Italy}
\address[2]{INFN, Sezione di Milano-Bicocca, Piazza della Scienza 3, I-20126 Milano, Italy}
\address[3]{Solar-Terrestrial Centre of Excellence - SIDC, Royal Observatory of Belgium, Avenue Circulaire 3, 1180, Uccle, Belgium} 
\address[4]{Center for mathematical Plasma Astrophysics, Department of Mathematics, KU Leuven, Celestijnenlaan 200B, 3001, Leuven, Belgium}
\address[5]{INAF, Osservatorio Astronomico di Cagliari, Via della Scienza 5, 09047 Selargius, Italy}
\address[6]{ASTRON $-$ the Netherlands Institute for Radio Astronomy, Oude Hoogeveensedijk 4, 7991 PD Dwingeloo, The Netherlands}

\received{\today}
\finalform{XXXXXX}
\accepted{XXXXXX}
\availableonline{XXXXXX}
\communicated{X. XXXXXX}

\begin{abstract}
In space weather studies and forecasting we employ magnetohydrodynamic (MHD) simulations which can provide rather accurate reconstruction of the solar wind dynamics and its evolution. However, all MHD simulations are restricted by the input data and the modelled solar wind characteristics need to be validated with different types of observations. That is very difficult, in particular for the solar wind characteristics close to the Sun, since the majority of in-situ observations are taken in the vicinity of the Earth. This is why all alternative methods for estimation of solar wind plasma characteristics are very important. In this study we utilise low radio frequency observations of pulsars to probe the total electron content along the line of sight. For the first time, we compare density estimates from pulsars with predictions from the 3D MHD modelling code; the EUropean Heliospheric FORecasting Information Asset (EUHFORIA). We find a very good correlation for the solar wind density along a given line of sight obtained by EUHFORIA and pulsar observations. We also demonstrate that the pulsar observations can be very useful not only for the model validation but also for understanding its limitations.      
\end{abstract}


\end{frontmatter}


\section{Introduction}\label{sec:intro}
The current decade has seen an optimistic outlook on the emergence of a commercial space age. This leap in human endeavour relies critically on an ever expanding network of satellites and other space based infrastructure that is constantly influenced by the space weather conditions -- the variable conditions in the heliosphere mainly driven by solar activity (see e.g. Bisi et al., this issue). Therefore, an accurate prediction of the space weather is of utmost importance. The aim of such predictions is to preserve not only Earth-based communication networks, but also the trillions of dollars of current and planned assets likely to populate near the Earth space \citep{ebh+17}. The local space weather conditions also influence our day to day lives, which depend increasingly on infrastructure such as energy distribution grids \citep[see e.g.,][ and references therein]{ehb+18, wkk+09}, logistics and supply chain networks \citep[][]{hapgood17, ffr16} and communications networks \citep[see e.g.,][]{bak04, lan07}, all of which can be affected by strong Geomagnetic storms. Thus the ability to accurately model and predict space weather events is an increasingly valuable tool.

During the last few decades, magneto-hydrodynamic (MHD) simulations have become indispensable in both scientific studies and operational forecasting of space weather conditions. Remarkable progress has been achieved in modelling the evolution and propagation of the solar wind (SW) and coronal mass ejections (CMEs) by employing 3D MHD models such as ENLIL \citep[][]{Odstrcil99, owens08, Taktakishvili09, Falkenberg10} and the EUropean Heliospheric FORecasting Information Asset \citep[EUHFORIA;][]{pp18, Scolini18, Verbeke2019, Scolini2021}. The recently developed, state-of-the-art space weather forecasting tool EUHFORIA \citep[][]{pp18} will be employed in this study. Although EUHFORIA models the evolution of CMEs \citep[e.g. see][]{Scolini18, Scolini19, Verbeke2019, Scolini20, Kilpua2019, Jebaraj20, Scolini2021, asv21} and the background SW \citep[][]{Hinterreiter19} considerably well, the validation of EUHFORIA's performance using different observations is necessary for further development of the model. 

With this study, we aim to test the modelling results of EUHFORIA, by comparing them with measurements of the free-electron column density of the SW that can be obtained from observations of radio pulsars. These rapidly rotating, highly magnetised neutron stars produce narrow beams of radiation at radio frequencies, approximately aligned with their magnetic poles, producing characteristic pulse-like signatures in ground based radio observations. These pulses are an excellent probe of the ionised media present along the line-of-sight (LOS) from the pulsar to an observer at Earth \citep[see e.g.,][ and references therein]{cou68,kcs+13,dvt19}. In particular, from pulsar observations we derive the free-electron column density of such media, also known as dispersion measure (DM). The \textit{pulsar timing} method \citep{Taylor1992, lk04} can provide estimates of the times-of-arrival (TOAs) of pulsar radiation to a precision of the order \SI{\sim 10}{\micro\second} to a few \SI{\sim 100}{\nano\second}. Combining the TOAs with low frequency observations, we can detect fluctuations in the DM at the level of \numrange{10}{1000} parts per million \citep[][]{tsc21}, and this allows us to decouple the interstellar and interplanetary components in the DM estimates \citep{tvs19,tsc21}. Using the interplanetary contribution to DM we build an independent probe of the SW density, which can be compared with the modelling results from EUHFORIA, and help in the fine tuning of the model. We note that this study is the first comparison of the SW densities from  pulsar observations and modelling results by EUHFORIA.

The article is organised as follows: in \Cref{sec:datamethod} we list the employed observations and report about the SW 3-D modelling given by EUHFORIA, and the derivation of pulsar-based electron column densities. In \Cref{sec:res} we present the results obtained by the two methods and their comparison. The summary and discussion are presented in \Cref{sec:disc}.

\section{Data sets and methods}\label{sec:datamethod}
In this study we make use of a number of different observations: (a) radio observations from the LOw Frequency ARray (LOFAR, \citealt{vh13}) telescope; (b) synoptic magnetograms from the Global Oscillation Network Group \citep[GONG;][]{GONGmagnetograms}; and (c) EUV observations from the SDO/AIA \citep[SDO;][]{SDO,Brueckner95}. We also make use of white light images from the Large Angle and Spectrometric COronagraph (LASCO) instrument on board the SOlar and Heliospheric Observatory (SOHO) spacecraft \citep{Brueckner95}. 

\subsection {Observations}
The \textit{radio observations} of pulsars were taken by the LOFAR telescope, presently the largest operating low-frequency radio interferometer. LOFAR operates in the frequency range from \SIrange{30}{240}{\mega\hertz} and consists of individual stations placed across Europe. Each station consists of arrays of two different antenna sets, covering the low (\SIrange{30}{80}{\mega\hertz}) and high (\SIrange{110}{240}{\mega\hertz}) frequency bands. While 38 stations are located in the Netherlands, a further 13 sub-arrays are international and can function as independent telescopes. Of those, the six international stations in Germany (referred to as DE601, DE602, DE603, DE604, DE605, and DE609) are operated by members of the German Long-Wavelength (GLOW) consortium (\url{https://www.glowconsortium.de/}) and are mainly used to carry out a long-term, high-cadence monitoring campaign of pulsars. For the work presented here, we make use of the publicly available dataset of \citet{dvt+20} based on GLOW observations. Additionally to the GLOW observations, we also employed a small number of observations taken with the central LOFAR stations. 

The second type of observations employed herein are the GONG magnetograms that are used as the main input to EUHFORIA model. These are synoptic maps in which data from observations are assembled over the course of a solar rotation (more information about synoptic magnetograms can be found at http://jsoc.stanford.edu/jsocwiki/SynopticMap). GONG magnetograms contain the magnetic flux density information necessary for the reconstruction of the 3D coronal magnetic field in EUHFORIA.

We also inspected EUV and white light observations. The first were obtained from the Atmospheric Imaging Assembly \cite[AIA,][]{Lemen2012} instrument on-board the Solar Dynamics Observatory \citep[SDO;][]{SDO}. The second (white light) observations were provided by the Large Angle and Spectroscopic Coronagraph \citep[\textit{LASCO};][]{Brueckner95} on board the Solar and Heliospheric Observatory \citep[SOHO;][]{SOHOmission} as well as by the Solar TErrestrial RElations Observatory Ahead \citep[STEREO A;][]{Kaiser08} for the time interval of available pulsar observations. Although the pulsar observations coincide with the low level of solar activity, we also check the existence of possible solar eruptions, e.g. CMEs or streamer blow outs. Such solar transients would strongly influence the plasma conditions and the density profiles which we obtained from both, radio observations and modelling. As the study focuses on SW densities, the observations including solar transients are to be avoided. To obtain these dates we make use of a modified version of the CMEchaser package \citep{stz20}. CMEchaser was originally created to search for CME occultations of background radio sources, and takes as input source coordinates in the celestial or galactic frames and locates the point at which the LOS to those coordinates will cross the helioprojective plane passing through the poles of Sun to then estimate the likelihood of a known CME encountering that point. Our modifications to CMEchaser allow us to recover the helioprojective coordinates of the point of crossing as a pair of angles, for each epoch.

\subsection{Pulsar selection}

In this study we estimate the SW contribution to the pulsar DM using sensitive low-frequency observations from the LOFAR telescope. LOFAR observations allow us to estimate the variability in the SW contribution as a function of the Solar elongation along the LOS. As the SW electron density varies as a function of the heliolatitude, we also constrain our pulsar selection such that the LOS lies relatively close to the Solar Equatorial plane. Additional reasons for this selection are the modelling constraints. The synoptic magnetograms which are the main input to EUHFORIA suffer from the strong projection effects in the regions close to the solar poles. Therefore EUHFORIA reliably models only regions close to the solar equator, i.g. the region covering \num{\pm 60} degrees. For some of the pulsar observations the LOS between the Earth and the pulsar can cross the unmodelled polar regions or the inner boundary of EUHFORIA. As a consequence the density obtained by EUHFORIA for the pulsars observed at the high heliolatitudes will be inaccurate. 

Since we use the method of precision pulsar timing to extract the DM time-series from which we compute the relative SW contribution, we selected a pulsar which was observed over a number of years, and whose astrometric and pulsar-intrinsic properties are well investigated \citep{vlh16,dvt+20,rsc+21}. This pulsar, J1022$+$1001, has a low ecliptic latitude of \num{-0.06 \pm 0.00}, which allows the LOS to directly cross through the Solar disk during conjunction in the period of August-September each year. Further, J1022$+$1001 was regularly observed by the German LOFAR stations (GLOW) since 2013 \citep{dvt+20}. These observations were used to build an accurate model for the contribution of the interstellar medium to the observed total DM allowing a more precise estimate of the SW contribution \citep{tsc21}.

\subsection {Methods}

The two main methods employed in this study are: 
\begin{itemize}
\item An accurate, epoch-wise estimate of the SW contribution to the pulsar DM estimate. 
\item A prediction of the SW properties from EUHFORIA, i.e., the variation of the SW in a 3D volume as a function of the Solar rotation. 
\end{itemize}

We present a summary of the employed methods below, details of which can be found in \cite{tsc21,dvt+20} and references therein for the pulsar DM estimation, and \citet{pp18, Hinterreiter19, Samara2021} for details on the EUHFORIA modelling. 

\begin{figure*}
\centering
\begin{subfigure}[b]{0.48\textwidth}
 \centering
 \includegraphics[width=\textwidth]{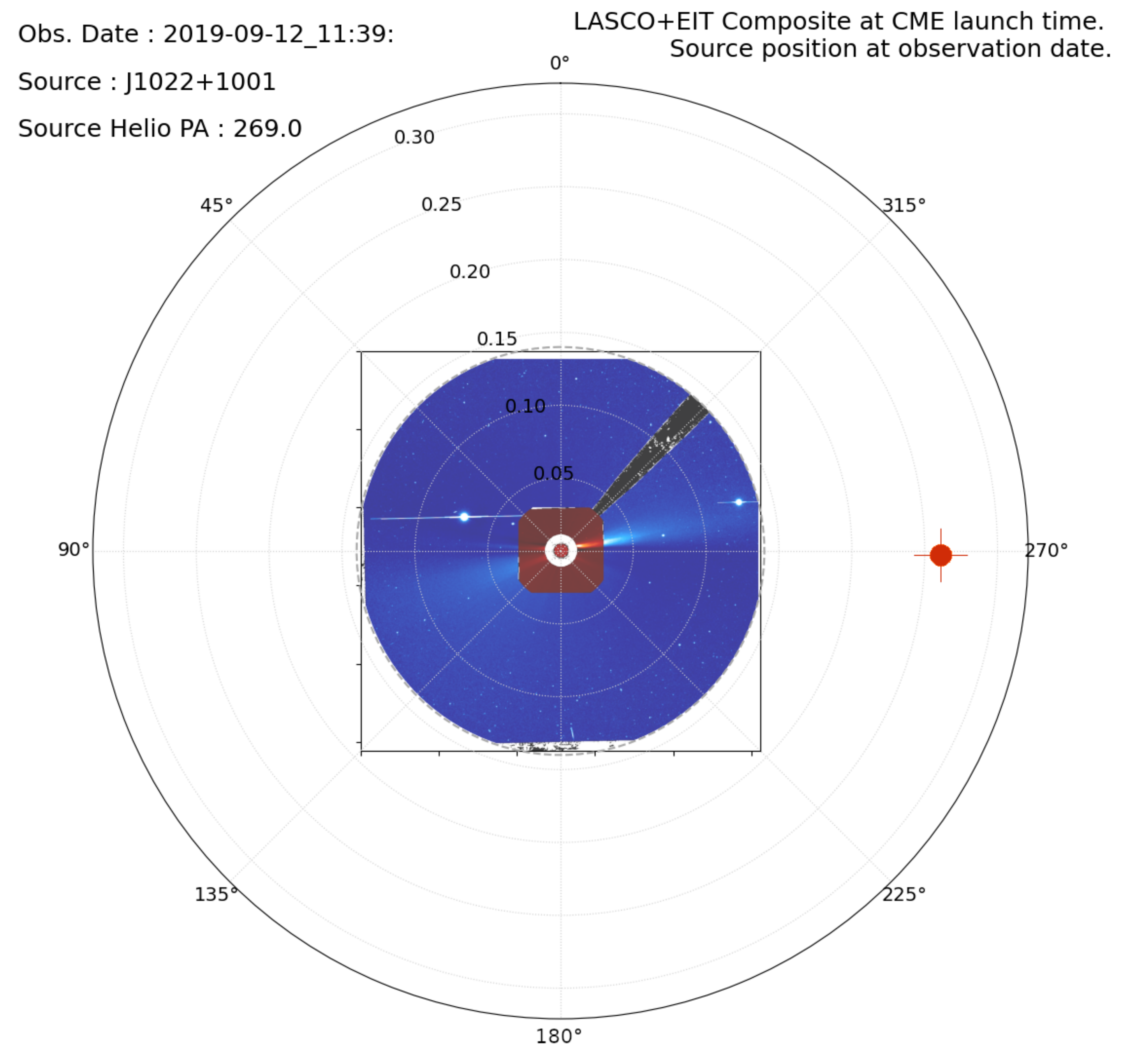}
 \caption{CMEchaser \protect\citep{stz20} plot; showing the relative position of the pulsar on the 12\textsuperscript{th} of September, 2019. Coordinates are the poistion angle and the elongation in \SI{}{\astronomicalunit}.}
 \label{fig:cmechaser_plot}
\end{subfigure}
\hfill
\begin{subfigure}[b]{0.48\textwidth}
 \centering
 \includegraphics[trim={2cm 0 0 0},width=1.05\textwidth, height=0.36\textheight]{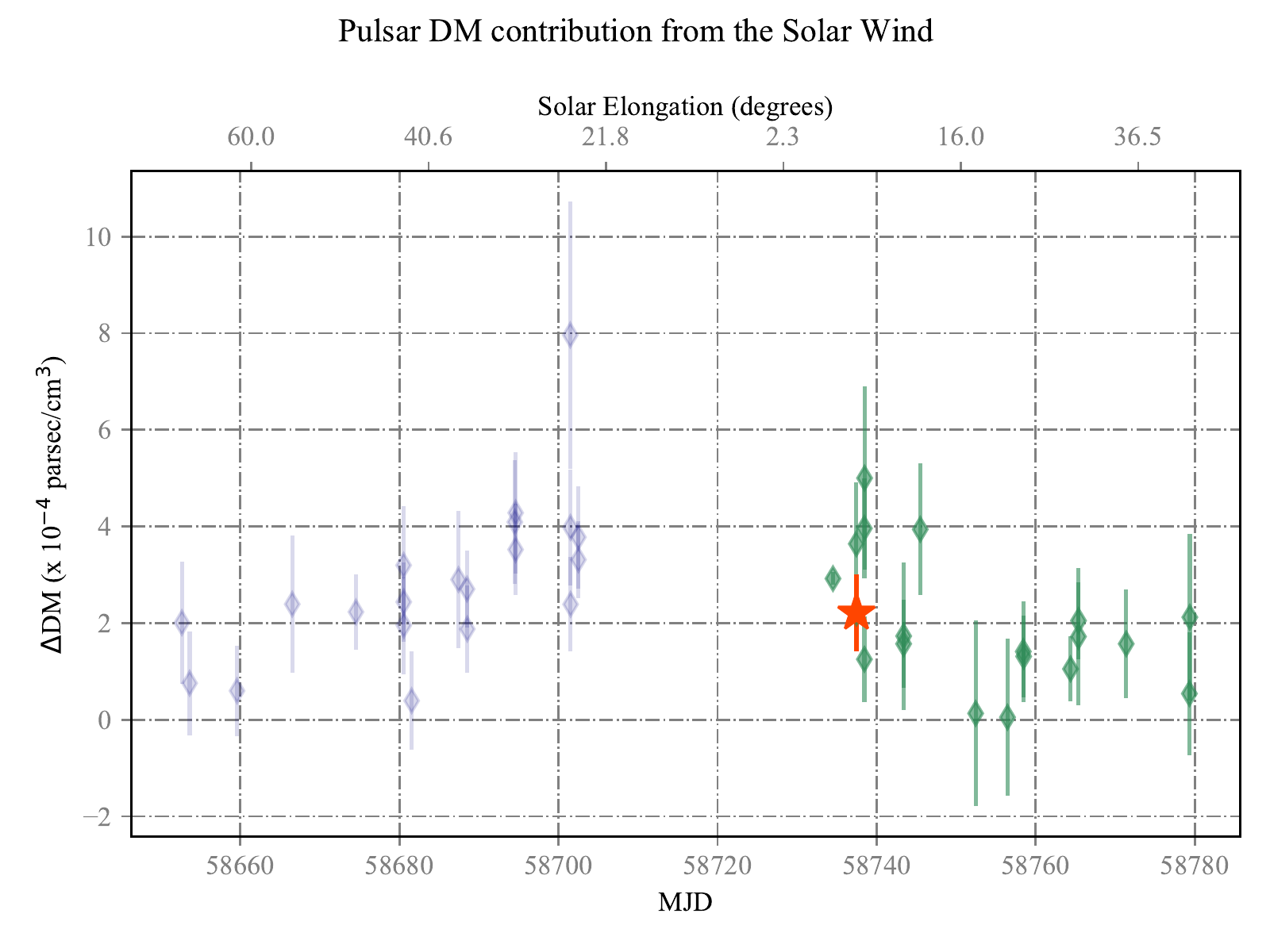}
 \caption{Estimates of the SW contribution to the pulsar DM, represented as a fluctuation around a mean DM. Note the increase in the excess as the solar (angular) separation decreases.}
 \label{fig:pulsar_estimates}
\end{subfigure}
\caption{Panel (a) shows the helio-projective position of the pulsar (red marker), overlaid on LASCO C2 and C3 images for the epoch of the radio observation marked with a red star in panel (b). Pulsar DM from the SW (diamonds with error bars), producing following the steps described in \cref{sec:datamethod}. The blue points represent the ingress of the LOS, while the green markers denote the egress. In this article we utilise only the points in the egress region. The red star marks the pulsar DM estimate for the 12\textsuperscript{th} of September, 2019. }
\label{fig:pulsardmsw}
\end{figure*}

\subsubsection{Estimation of the Pulsar Dispersion Measures from radio observations}\label{sec:dm}

The technique we employed to estimate a DM value is the aforementioned \textit{pulsar timing} method. This method employs a linearised least squares approach \citep{tay92, ehm06, hem06} to provide a \textit{timing model}, i.e., a set of parameters which characterise the pulsar and the medium along the LOS towards the pulsar. Since the ambient medium is in our case partially ionised, radio frequencies propagating in these media experience a frequency dependent group delay. This dispersive effect is characterised by the `dispersion measure', $ \mathrm{DM} \equiv \int_{0}^{L} n_{e}dl $ where $n_e$ is the electron density and $L$ is the distance to the pulsar. The DM can be separated into two components, the first arising from the extended propagation through the tenuous ionised interstellar medium, and the second due to propagation in the SW. 
 
An updated DM time series for PSR~J1022$+$1001 was derived following the same steps as in \citet{tsc21}, starting from the unfiltered TOAs provided by \citet{dvt+20}. Although \citet{dvt+20} have already applied radio frequency interference (RFI) removal to the observations, we first re-analysed the set of 10 frequency-resolved ToAs per observation for bad data points due to low level RFI or pulsar intrinsic effects. Each ToA refers to a frequency channel of \SI{\sim 2.3}{\mega\hertz} bandwidth. Outliers in the ToAs were identified using Huber regression \citep{hub64}, as data lying further than three times the median absolute deviation from a robust fit. The underlying model for this fit followed a $f^{-2}$ dependence where $f$ is the frequency associated with each TOA. The resulting set of epoch-linked TOAs was again fitted using the \textsc{Tempo2} software package and the accompanying timing model, which was modified to remove any model for temporal DM variations. To produce the DM time-series for PSR~J1022+1001, we take the timing model updated in the last step and now fit for only the DM parameter at each observing epoch. It is important to note that during the timing procedure, we ensured that the default SW modelling schemes of \textsc{Tempo2} were not employed. Thus each point in the resulting DM time-series represented an epoch-wise estimate of the total DM delay, which must be separated into the contribution from the IISM and the SW.\\

To remove the influence of the IISM, for each pulsar we followed the method presented in \citet{tsc21}, which we describe briefly here. 

The multi-year DM time-series was first divided into 460-day long segments centred on the Solar conjunction of PSR~J1022+1001 and segments shorter than this time-span were excluded. Then, the DM time-series in each segment was modelled in a Bayesian framework as the sum of contributions from a spherically-symmetric SW model and a third order polynomial, representing the IISM contribution chosen following explanations in \citet{tsc21}. The spherical SW model was parametrised as a function of the electron density at \SI{1}{\astronomicalunit}; $\rm{A_{\SI{}{\astronomicalunit}}}$ and the Solar elongation $\rho$ of the pulsar:
\begin{equation}\label{eq:dmsunsph}
DM_{sw} = 4.85 \times 10^{-6} \mathrm{A_{\SI{}{\astronomicalunit}}} \frac{\rho}{\sin\rho}~\mathrm{pc~cm^{-3}}.
\end{equation}
The model accounts for the systematic errors in the DM estimates using an additional parameter, which is summed in quadrature with the DM uncertainties in the likelihood function. Parameter estimation is performed using the Markov Chain Monte Carlo method \citep[implemented via the {\it emcee} package;][]{fhlg13}. To minimise observer bias injection, flat priors were used for all free parameters but the amplitude of the SW model and the uncertainties correction parameter were also constrained to be positive. Finally, the recovered polynomial model for the IISM contribution was removed from the DM time series to produce the SW contribution estimate. \Cref{fig:pulsar_estimates} shows the obtained time-series for the SW contribution to the DM for PSR~J1022+1001 from the 12\textsuperscript{th} of June to 29\textsuperscript{th} of September, 2019. Following the reasoning at the start of this section, we utilise only the SW DM estimates from September and October, 2019 to carry out the comparison.

\subsubsection{The 3D MHD modelling with EUHFORIA}\label{sec:mhdmodelling}

\begin{figure}
\centering
\includegraphics[scale=0.92]{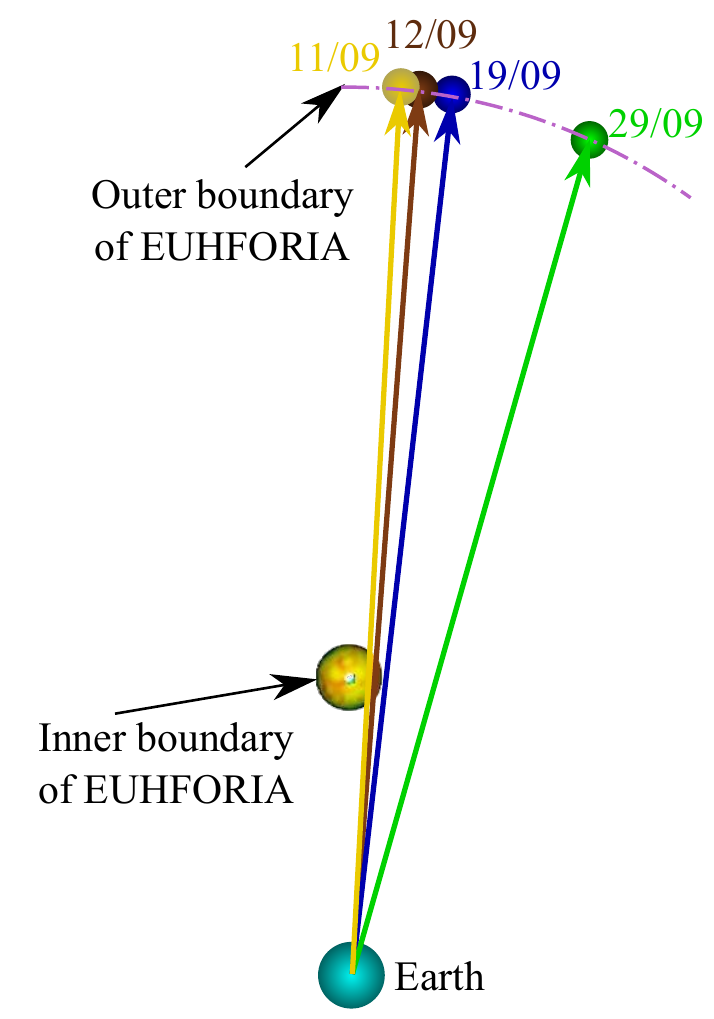}
\caption{Relative positions of Earth, the inner boundary of EUHFORIA's simulation and the point at which the LOS crosses the simulated volume boundary, projected of the Solar Equatorial plane. Although it is almost indistinguishable at this scale, the LOS on 11/08/2019 passes through the inner boundary of the EUHFORIA simulation volume, imposed by the latitudinal limits on the magnetograms. }
\label{fig:positions}
\end{figure}

EUHFORIA \citep[][]{pp18} is a 3D MHD space weather forecasting tool for modelling the SW and CMEs within the inner heliosphere.
The modelling domain of EUHFORIA is separated into two parts: the coronal part, extending from the solar surface up to \SI{0.1}{\astronomicalunit} (so called inner boundary of EUHFORIA), and the heliospheric part, covering distances from \SIrange{0.1}{2}{\astronomicalunit} \citep{pp18}.

The main input to EUHFORIA is a solar photospheric magnetogram. Although the model can accept different input magnetograms, the most frequently employed ones are from GONG. After the magnetogram insertion, the semi-empirical Wang-Sheeley-Arge \citep[WSA;][]{arge03, Arge04} model is used in combination with the potential-field source-surface \citep[PFSS;][]{altschuler69, Wiegelmann2017}, and the Schatten Current Sheet \citep[SCS;][]{schatten69} models, to obtain the boundary conditions at \SI{0.1}{\astronomicalunit}. These boundary conditions are necessary to initiate the heliospheric counterpart beyond \SI{0.1}{\astronomicalunit}. 

\begin{figure*}
\centering
\includegraphics[]{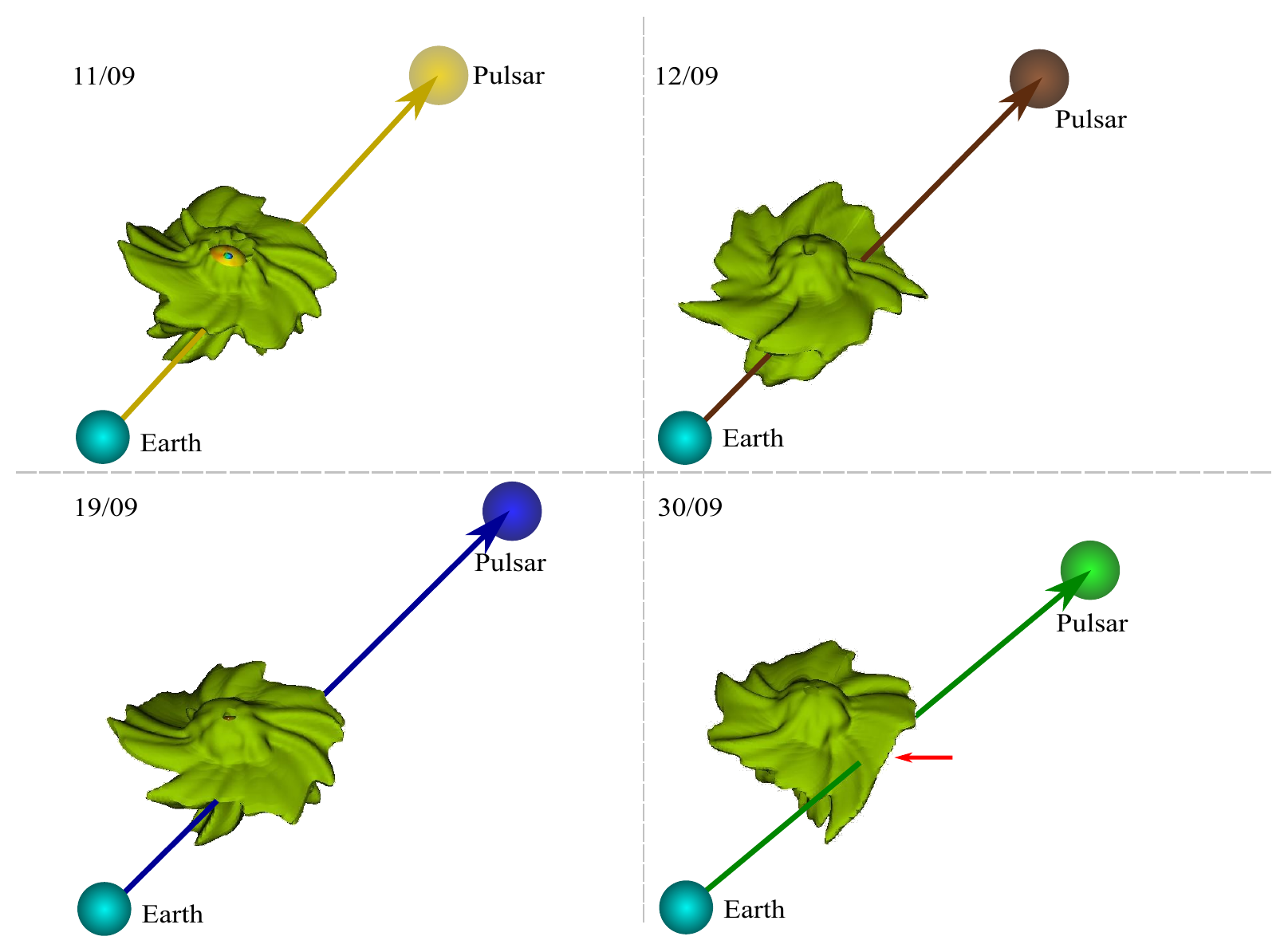}
\caption{Relative positions of the Earth (cyan sphere), the point at which the LOS punctures EUHFORIA's simulation boundary (monochromatic spheres, marked ``Pulsar") and the inner boundary of the EUHFORIA simulation (multicolor sphere). Also shown in light green is the iso-surface of density n = 100 particles/cm\textsuperscript{2}. Note the detailed iso-surfaces of density and their 3D properties.}
\label{fig:3dpositions}
\end{figure*}

\section{Results}\label{sec:res}

\subsection{Modelling results}\label{sec:res1}
\begin{figure*}
\centering
\includegraphics[scale=0.86]{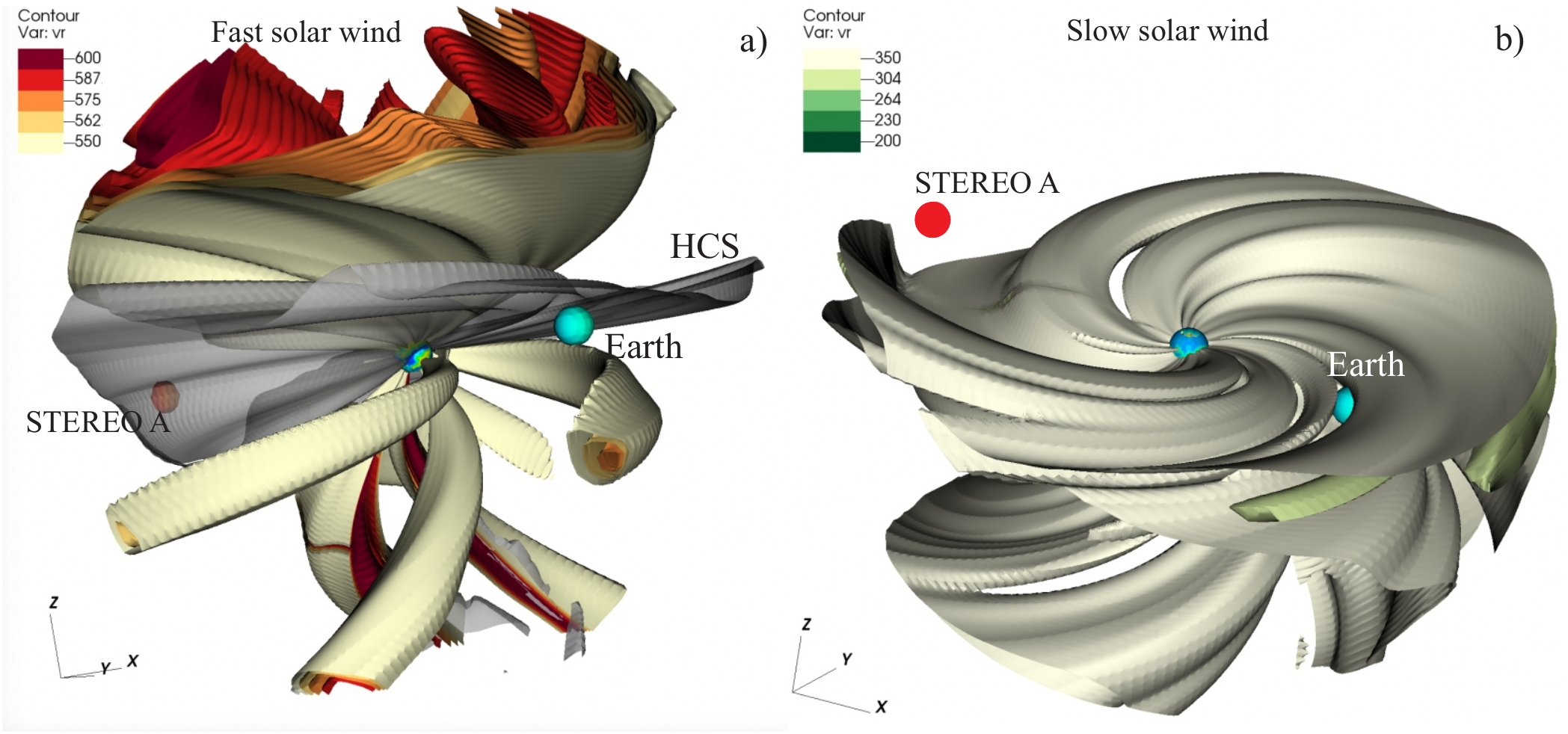}
\caption{The SW velocity modelled by EUHFORIA on 19/09/2019. The velocity is represented with the colour coded iso-surfaces. The blue and red spheres present Earth and STEREO A, respectively. The colourful sphere in the centre indicates the inner boundary of EUHFORIA at \SI{0.1}{\astronomicalunit} and the grey iso-surface is the heliospheric current sheet (HCS), $B_r$ = 0. Panel a) shows the iso-surfaces of the fast SW originating mostly from the polar regions and panel b) shows the iso-surfaces of the slow SW observed mostly in the equatorial regions.}
\label{fig:sw}
\end{figure*}

\begin{figure}
\centering
\includegraphics[scale=0.7]{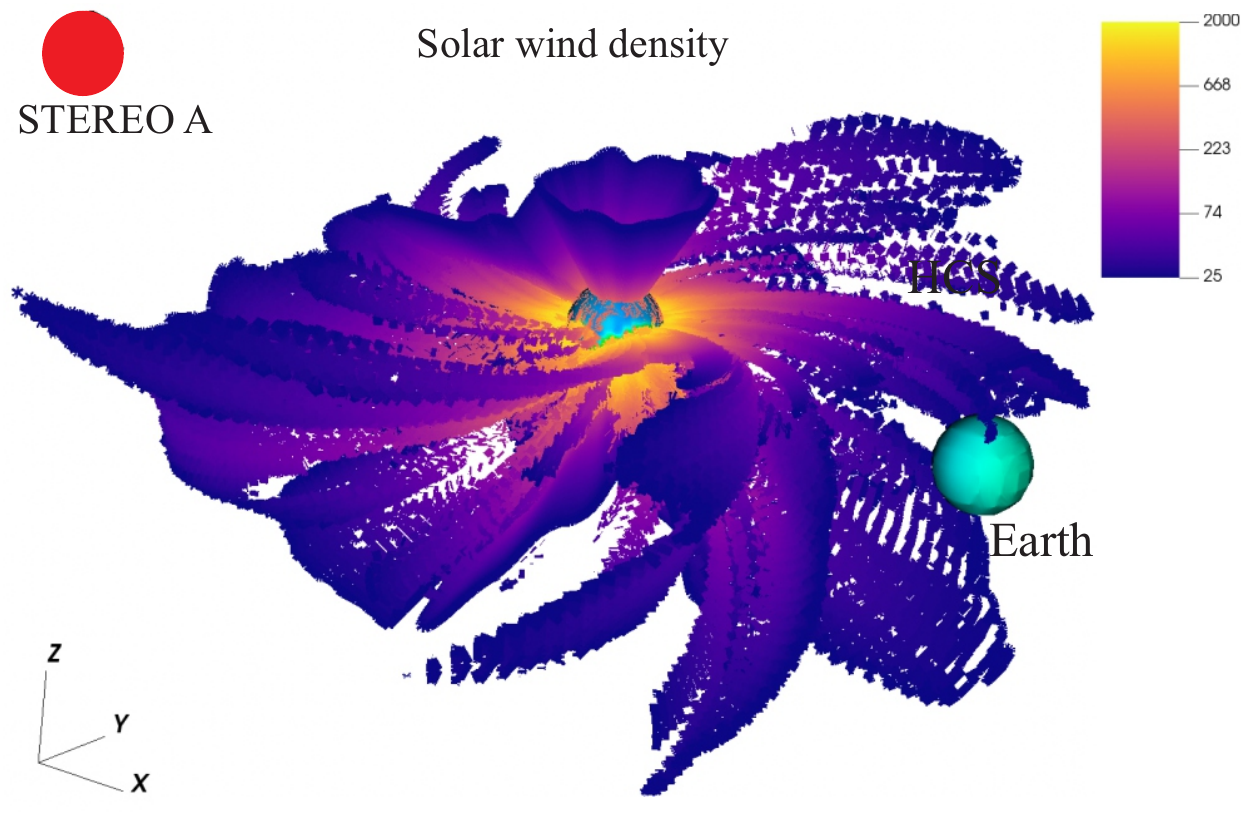}
\caption{The SW density modelled by EUHFORIA on 19/09/2019. The blue and red spheres present Earth and STEREO A, respectively. The colourful sphere in the centre indicates the inner boundary of EUHFORIA at \SI{0.1}{\astronomicalunit}. Density is represented by the colour coded iso-surface with the lighter, orange and yellow regions showing the high density and the darker, violet lower density.}
\label{fig:n}
\end{figure}

For the selected dates of pulsar observations in \Cref{tab:euhvspsr}, we modelled the SW with EUHFORIA by employing GONG magnetograms as input \citep[][]{GONGmagnetograms}. 
The input magnetograms were taken  at  12:00 UT on the day of pulsar observations. \Cref{fig:sw} and \Cref{fig:n} show the example of the modelled SW in the 3D space on the 19\textsuperscript{th} of September, 2019.  

\Cref{fig:sw}a shows that EUHFORIA models the fast SW originating from regions closer to the poles while  \Cref{fig:sw}b shows that the slow SW originates mostly from equatorial regions. the density of the slow SW is generally considered to be higher than that of the fast SW. Therefore, as we consider pulsars for which the LOS path crosses the equatorial regions of the slow SW (\Cref{fig:sw}b), a significant contribution of the slow SW densities to the density estimated from pulsars is expected. \Cref{fig:n} shows a wire frame (of the iso-surface) plot of the SW density in the 3D space modelled by EUHFORIA. The high densities (yellow and orange colours) are mostly found in the regions closer to the Sun. As we move to larger distances from the Sun the SW density decreases (violet colour). A similar presentation to that of the SW velocity (\Cref{fig:sw}) is not possible for density as density decreases exponentially in the anti-sunward direction and therefore a regular iso-surface plot would only show the lowest SW density values (similar to presentation in \Cref{fig:3dpositions}).

\subsection{Comparison of density obtained from pulsar observations and from modelling with EUHFORIA}\label{sec:res}

\begin{figure}
 \centering
\includegraphics[trim={0 0 0 .4cm},clip,height=0.9\textheight, width=\linewidth]{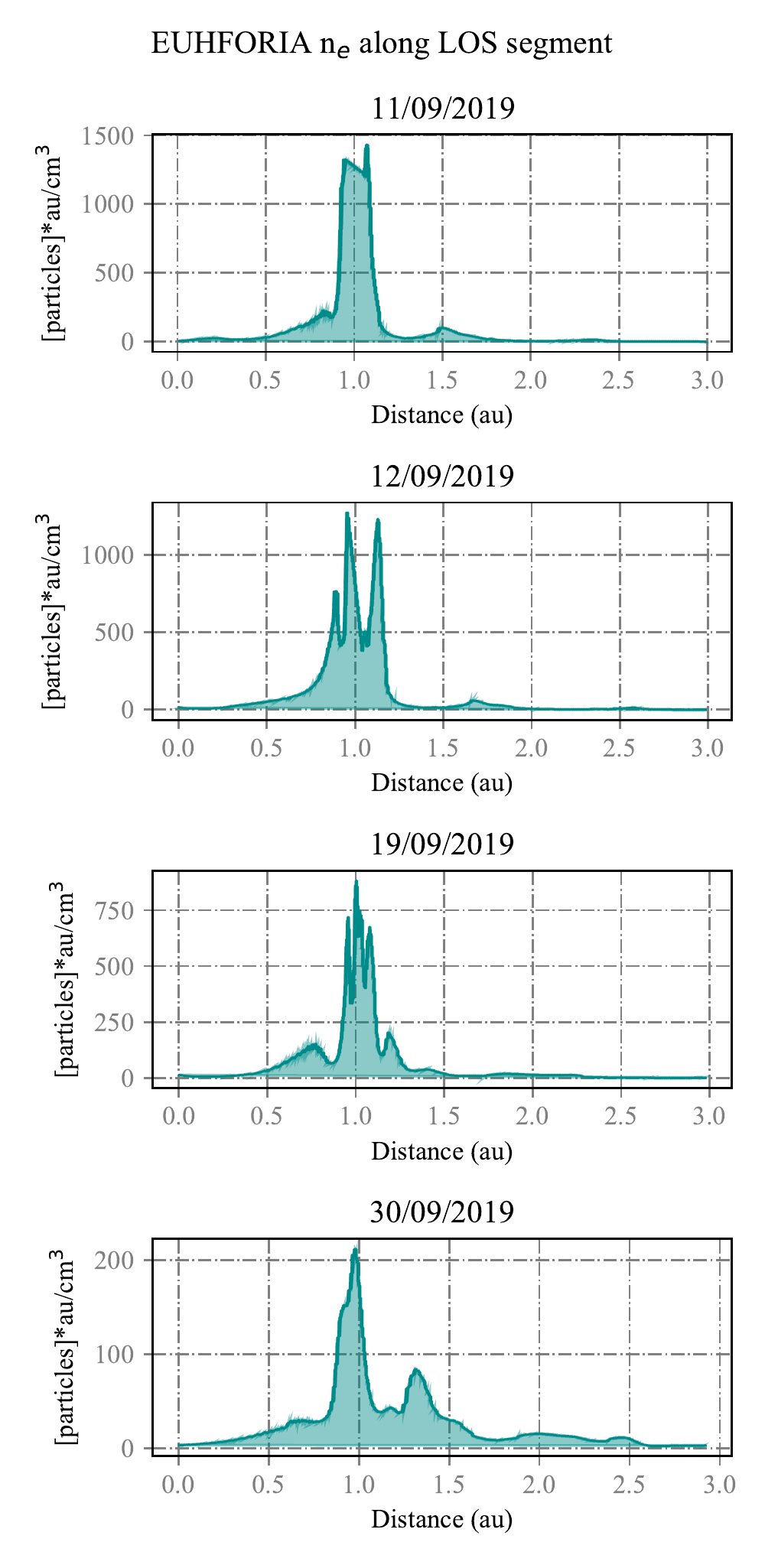}
\caption{Electron density along the LOS segment within the EUHFORIA volume, as a function of the distance from Earth. In this case, the Sun lies at \SI{\sim 1}{\astronomicalunit}.}
\label{fig:curves}
\end{figure}

The 3D outputs of EUHFORIA allow us to compare the modelled SW density across a wide range of LOS orientations, and for different pulsar positions. \Cref{fig:positions} shows the relative positions of the observed pulsars, Earth and the inner boundary of EUHFORIA (the monochromatic spheres on the curved line, the cyan and the multicolored spheres, respectively). The monochromatic spheres are the points at which the LOS crossed through the simulation boundary of EUHFORIA. The multi-colour sphere represents the EUHFORIA's inner boundary, i.e. the modelled SW characteristics at \SI{0.1}{\astronomicalunit}. The corresponding set of modelling results is displayed in \Cref{fig:3dpositions}, showing the iso-density contour of \SI{100}{[particles]\per\cm\cubed} in light green. Here the Earth is shown again in cyan and the point at which the LOS crossed beyond the simulation volume presented as monochromatic spheres. After the SW modelling an additional integration step was performed to convert the 3D density values, such as those presented in \Cref{fig:n}, into a column density. To do this, we first produce curves for the point-wise electron density along the LOS segment lying within the simulation volume. The four curves for the corresponding dates are shown in \Cref{fig:curves}. To obtain the column averaged electron density, we fit a single Gaussian component along with a constant offset to each of these curves and take the value at the full-width half-maximum as the mean value. We propagate the fit errors to recover a statistical error on these mean values. Since EUHFORIA does not provide an error on the density values, we estimate an error range by comparing the density estimates for the same LOS segment obtained using pairs of successive magnetograms separated by 24h hours. We find that this shows an almost identical offset of \SI{\sim 10}{\per\cm\cubed}. We add this range in quadrature to the statistical errors and the values are presented in the fourth column of \Cref{tab:euhvspsr}.

\begin{figure}
\centering
\includegraphics[trim={0.5cm 0.5cm 0 0},clip,scale=0.58]{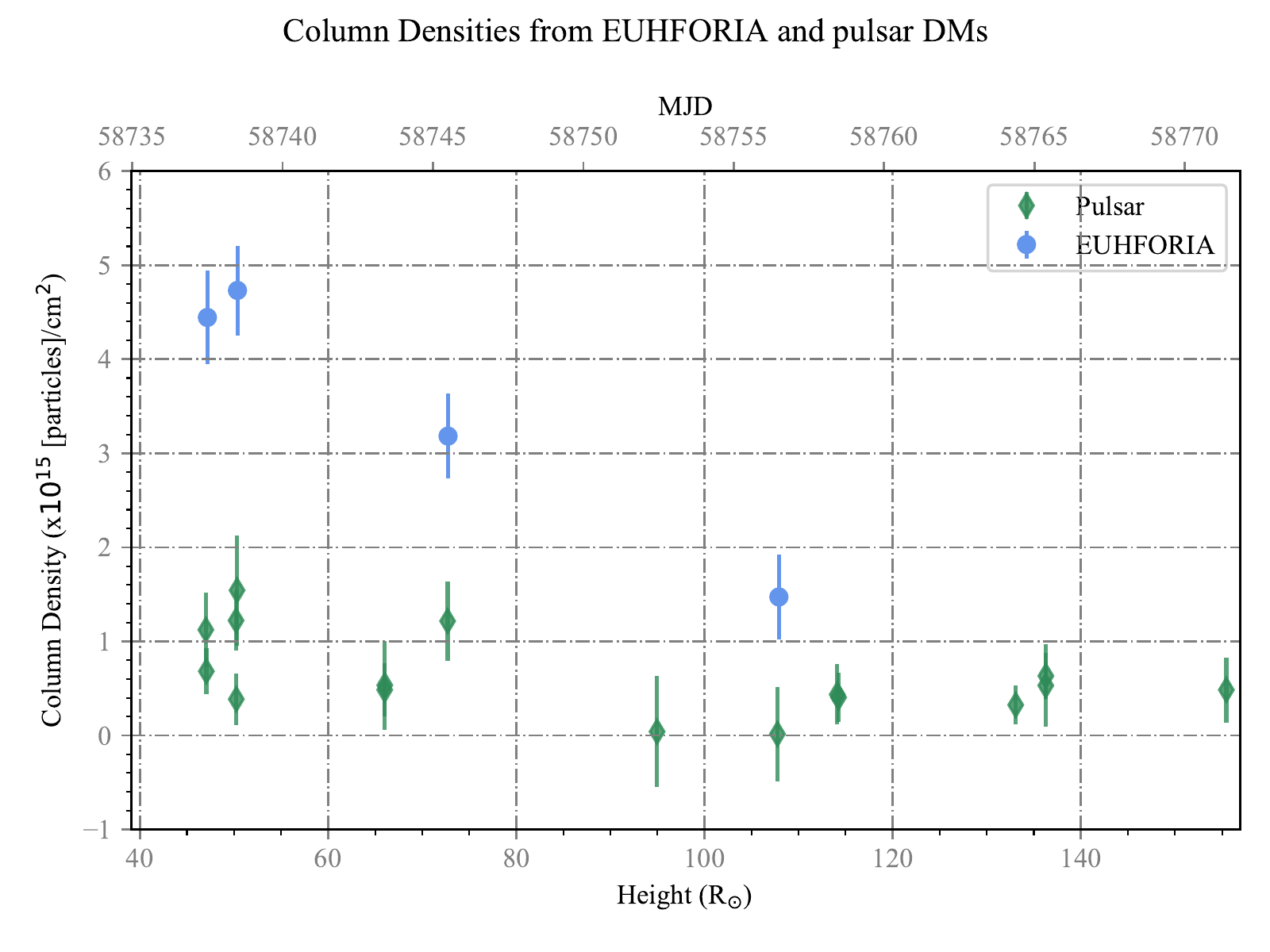}
\caption{Comparison of EUHFORIA predictions (filled blue circles) with pulsar DM estimates (green markers with errorbars) of the SW contribution. Here height denotes the perpendicular distance to the LOS from the Solar surface, on the plane passing through the center of the Sun and perpendicular to the Sun-Earth vector.}
\label{fig:euhvspsrdm}
\end{figure}

\begin{figure}
\centering
\includegraphics[trim={0.5cm 0.5cm 0.5cm 0},clip,scale=0.58]{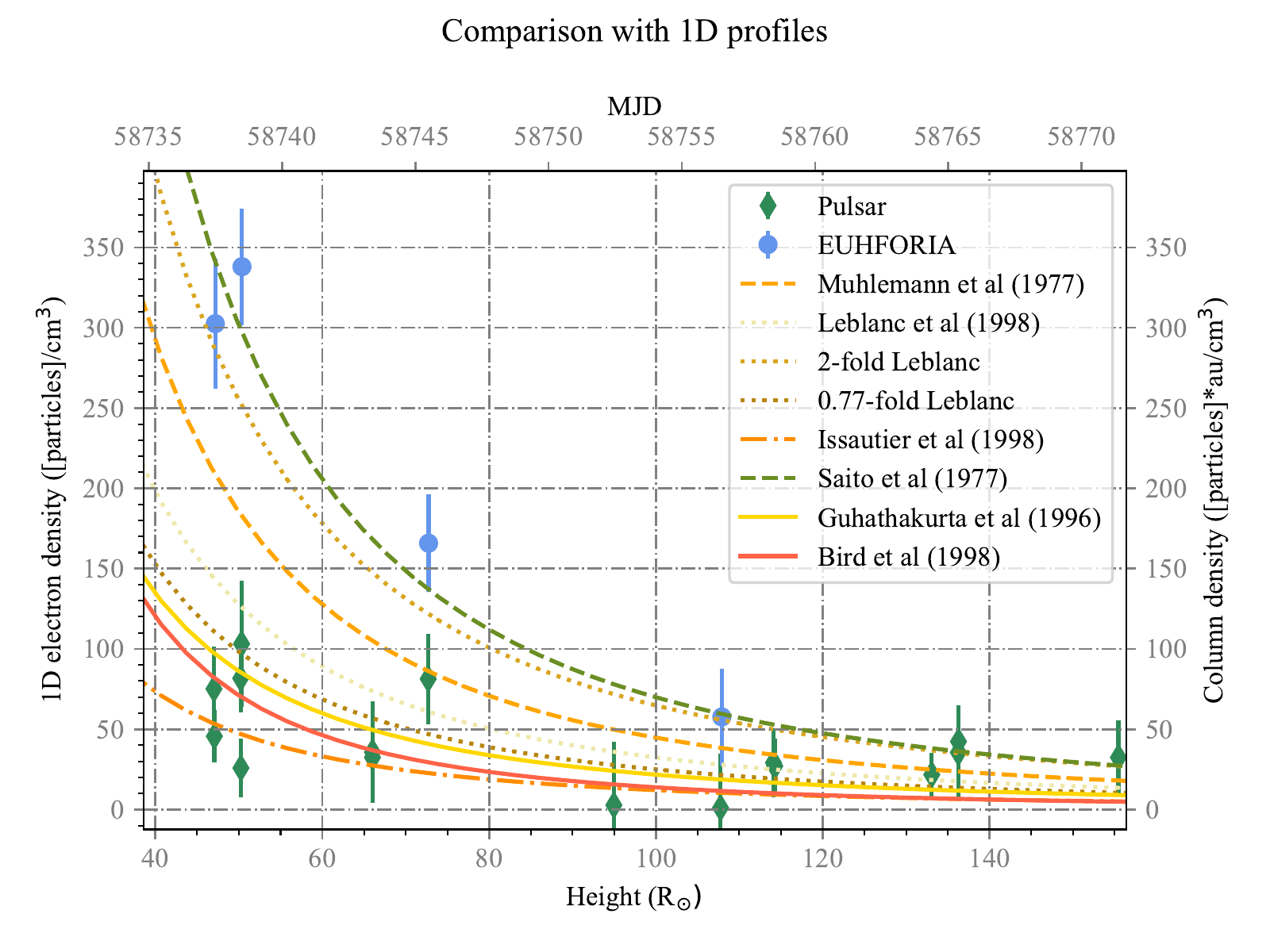}
\caption{Comparison of estimated column densities from EUHFORIA predictions (blue markers with errorbars) and pulsar DM estimates (green markers with errorbars) of the SW contribution. Also shown are predictions from 1D models of the SW density for the maximum contribution, i.e., at the point where the LOS crosses the plane passing through the Sun's poles. Note that the 1D profiles and the pulsar and EUHFORIA estimates are not identical quantities, as described in the text.}
\label{fig:euhvspsr}
\end{figure}

We also convert the SW contribution to the pulsar DM estimation using the steps mention in \Cref{sec:datamethod}, from the customary pulsar units of \si{\parsec\per\cm\cubed} to \si{\per\cm\squared}. Since a few additional days of pulsar observations, besides those listed in \Cref{tab:euhvspsr} were available, and in order to obtain an idea of the dynamics of the estimated DM values, we also plot them in \Cref{fig:euhvspsrdm,fig:euhvspsr} as a function of the height (in Solar radii) of the LOS above the solar surface, on the plane passing through the centre of the Sun and perpendicular to the LOS.  

The comparison between the pulsar-derived column density values and density obtained from the EUHFORIA simulations is shown in \Cref{fig:euhvspsrdm}. In that figure, the values that were obtained from the simulations for the first three days under consideration are somewhat larger. However, this behaviour can be explained by inspecting the \Cref{fig:3dpositions}. For the first three dates an extended portion of the LOS passes through the light green iso-density surface modelled by EUHFORIA. However on 30/09/2019, we can observe a dip in the  iso-density surface along the LOS towards the pulsar, indicated by the red arrow. Thus the LOS crosses a significantly smaller portion of the slow and dense SW than in a case of first three dates. As a result the modelled density on 30/09/2019 is somewhat smaller than for the previous three dates and very similar to that obtained from pulsar observations (\Cref{fig:euhvspsrdm} and \Cref{fig:euhvspsr}). 

The \Cref{fig:euhvspsr}, in which we compare the average electron densities obtained from the pulsar DM measurements and the EUHFORIA simulation results, indicates that the obtained values for the first three dates are also not extreme and can be expected at these height when accounting for dynamical evolution of the SW and its associated magnetic field structures. Clearly, the column densities obtained by the two methods, but now converted to units of \SI{}{\astronomicalunit\per\cm\cubed} in \cref{fig:euhvspsr} are lie within the range predicted by several 1D coronal electron density models. These models, which are frequently employed in studies of solar radio emission, are based on: (a) Mariner 6 and 7 flyby data \citep{mea77}; (b) polarised brightness images from the Skylab coronagraph \citep{spm77}; (c) type III radio bursts observations by Wind \citep{ldb98}; and (f) Ulysses high latitude fly-by data \citep{imm+98}. We also note that the 1D models provide coronal density estimation only at one point in space and time. Therefore we can consider them as a lower limit for the values that should be expected from the pulsar observations and modelling with EUHFORIA. We first discuss the comparison of the 1D models and the density obtained from pulsar observation. The differences in the density modelled by EUHFORIA, for the four considered dates, will be further elaborated in the \Cref{sec:disc}. 

The \citet{mea77} model shows higher values than the density obtained from pulsar observations. That is probably due to the point that the 1D model was obtained during the solar maximum, while our study is considering observations during the solar minimum. The \citet{imm+98} model, based on \textit{in situ} measurements of the SW with the Ulysses satellite, shows the best agreement with density obtained from pulsars observation. We note that the Ulysses dataset is best applicable at heliolatitudes $>40$ degrees. We also fit models from \citet{bpe+96} and \citet{mfs96} to the pulsar column density estimates, which lead to recovered models of the form:
\begin{equation}
N_e{(l)} = 89\times10^4\ \bigg(\frac{l}{R_\odot}\bigg)^{-2.4}\ \ \mathrm{cm^{-3}}\ \ .
 \end{equation}
 \begin{equation}
N_e{(l)} =10^7\bigg(\frac{l}{\mathrm{R}_\odot}\bigg)^{-4} +22\times10^4\bigg(\frac{l}{\mathrm{R}_\odot}\bigg)^{-2}\ \ \mathrm{cm^{-3}}
 \end{equation}
where $l$ is the distance from the centre of the Sun, in units of Solar radius, R$_{\odot}$ and we have ignored any heliolatitude dependence.

The \citet{ldb98} model, based on observations of the type III radio bursts during a solar minimum, provides slightly larger values than our estimation from the pulsar observations. Taking into account that this model was developed employing observations of radio bursts propagating along the open magnetic field lines and mapping the density along particular magnetic configurations, the agreement presented in \Cref{fig:euhvspsr} is good. The \citet{ldb98} model is often used with scaling factors, such that during the high level of the solar activity the 2-fold \citet{ldb98} model is used. In a case of large eruptive events, an even larger scaling factor is sometimes employed. The best fit with density obtained by pulsars is with 0.7-fold Leblanc. 

\section{Discussion and conclusions}\label{sec:disc}

This study presents, for the first time, comparisons of the SW densities obtained employing two very different methods; those estimated from pulsar observations and those from modelling by the MHD model EUHFORIA. 

\Cref{fig:euhvspsrdm} presents our results, i.e. the comparison of the column density obtained from the EUHFORIA runs and from the pulsar observations. We found a good correspondence between the column density obtained by the two very different methods. However the agreement is somewhat worse than that when the column density obtained from pulsar observations was compared with  interplanetary scintillation (IPS) measurements (see Tiburzi et al., this issue). This can result from a few different factors. First, in the study by Tiburzi et al. (this issue), the column density obtained from observations of pulsars is compared with that from IPS observations, i.e. only observations are employed in that study. In our study, the results obtained from pulsar observations are compared with the modelling results which necessitates different constraints on the comparison. 
The second reason for the somewhat larger variance in the results comes purely from the modelling. Namely, the only input to EUHFORIA, in a case of SW modelling, is the magnetogram. Magnetograms taken by different instruments and different observatories can differ significantly \citep[][]{Riley2014} and as a result, employing magnetograms from different sources as an input to the same MHD model will induce somewhat different results.

However, we will elaborate further on the difference in the column density obtained in modelling results by EUHFORIA for the four considered dates. \Cref{fig:euhvspsrdm} and  \Cref{fig:euhvspsr} show that the largest differences between the column density obtained from pulsar observations and the one modelled by EUHFORIA are obtained for the first three considered dates. On the other hand, the modelled column density for last date is in excellent correspondence with the one obtained from the pulsar observation. In order to try to understand why such different column densities were modelled by EUHFORIA we inspected the EUV images of the Sun which show clearly the existence of coronal holes, the source of the fast SW. We inspected EUV images one to three days before the pulsar observations. This time window was selected in order to account for the time needed for the SW or CME to travel across the distances at which we observe pulsars. Here, we considered the average SW velocity to be \SI{400}{\km\per\second} and \SI{1000}{\km\per\second} as the average CME velocity. 
The \SI{193}{\angstrom} EUV observations from the SDO/AIA instrument showed that in case of the first three dates, very patchy unipolar regions were observed on the solar disc. Although these were very irregular and structured regions the CHIMERA software \citep[][]{CHIMERA} isolated them as a coronal holes. As such these regions are the source of a rather fast SW, but probably not more than \SIrange[range-phrase=\textup{--}]{450}{500}{\km\per\second}, which is characterised by a lower density than the slow SW. The presence of a faster and less dense SW was thus recovered in the column density obtained from pulsar observations but not in the one modelled by EUHFORIA. Namely, as mentioned earlier, the first limitation in the SW modelling is induced by the employed magnetogram and the PFSS extrapolation. EUHFORIA, and probably any other MHD model that has requires at its seed input magnetogram input and PFSS, cannot model the SW originating from such patchy and structured coronal holes. Therefore, the modelling results will show the presence of predominantly dense and slow SW.

On the other hand, in the case of the last studied date, for which the correlation of obtained densities was very good, we observe two quite well defined, elongated equatorial coronal holes in the \SI{193}{\angstrom} observations. These two coronal holes are the source of the fast SW which is now modelled better by EUHFORIA than the SW from the patchy coronal holes of the first three dates. Since the fast SW is also less dense than the slow wind, the SW contribution to the column density modelled by EUHFORIA is now smaller and the values are very close to the one obtained from pulsar observations.

\begin{table}
\centering
\caption{Particle column density as a function of date\label{tab:euhvspsr}. Estimates from pulsar DM measures have been averaged to produce a mean for each observation date. The EUHFORIA values presented here are averaged following a Gaussian fitting procedure, see text.} 
\begin{tabular}{|l|l|l|l|l|}
\hline
Obs Date & MJD & Height&\multicolumn{2}{c|}{Column density (\SI{}{\astronomicalunit\per\cm\cubed})}\\
 \cline{4-5}
 & &(R$_{\odot}$)&EUHFORIA & Pulsar\\
\hline
11/09/2019 & 58737 & 46.92 &303(27) &144(37)\\
\hline
12/09/2019 & 58738 & 49.54 &338(21) &167(38)\\
\hline
19/09/2019 & 58745 & 67.86 &166(6)  &193(66) \\
\hline
30/09/2019 & 58756 & 96.64 &58(6)   &3(40) \\
\hline
\end{tabular}
\end{table}

Thus, the two most important outcomes of this study are: 
a) the trends of the density as a function of the distance from the Sun, obtained from two very different methods (pulsar observations and MHD modelling) are quite well correlated and within the level of densities obtained by other methods;
b) in case of SW originating from patchy and small coronal holes the density change is better mapped by the pulsar observations than by the MHD model. This results mostly from a large dependence of the modelling results on the initial magnetogram input and the employed PFSS extrapolation.

Our findings are not only promising for space-weather applications, but also for pulsar-based experiments such as Pulsar Timing Arrays (PTAs, see e.g. \citealt{tib18}), which aim to detect low-frequency gravitational waves by identifying corresponding perturbations in pulsar signals. In PTAs, SW dispersion can compromise the success of the experiment \citep{thk16}, therefore it has to be modelled and subtracted. The most realistic SW model available for PTAs was presented by \citet{yhc07}. Here, the SW is approximated as a two phase medium made up of a slow, dense component and a faster but rarefied one, whose locations are identified through the modelling results of EUHFORIA. The LoS to the pulsar is then divided into segments, whose DM contribution is evaluated depending on the SW component affecting them, and their radial distance to the Sun. The expected DM is obtained by summing the ensamble of the contributions along the same LoS. While this approach was found sub-optimal by \citet{tvs19}, the tests carried out so far with EUHFORIA open up an avenue for developing more accurate and realistic SW model for PTA purposes.

Our results show that inputs from pulsar measurements can be used to test model predictions from EUHFORIA. This early demonstration also validates the effectiveness of the Bayesian modelling \citep{tsc21} to separate the pulsar DM into contributions from the IISM and that from the SW. 
The presented experiment draws a parallel with the successful cross-validation between pulsar and interplanetary scintillation measurements reported by Tiburzi et al. (this issue). These results show the potential for the inclusion of radio-pulsar data in space weather studies.

\section{Acknowledgments}
GS acknowledges financial support provided under the European Union’s H2020 ERC Consolidator Grant "Binary Massive Black Hole Astrophysics" (B Massive, Grant Agreement: 818691). This work is part of the research program Soltrack with project number 016.Veni.192.086 (recipient C. Tiburzi), which is partly financed by the Dutch Research Council (NWO). I.C.J. and E.S were supported by a PhD grant awarded by the Royal Observatory of Belgium. I.C.J and J.M. acknowledge funding by the  BRAIN-be project SWiM (Solar Wind Modelling with EUHFORIA for the new heliospheric missions). This paper is partially based on data obtained with: LC12\_019.

\bibliographystyle{model5-names}
\biboptions{authoryear}
\bibliography{references,journals}

\end{document}